# 2008 LC18: a potentially unstable Neptune Trojan

Horner, J.[1], Lykawka, P. S.[2], Bannister, M. T.[3] & Francis, P.[3]

[1] Department of Astrophysics, School of Physics, University of New South Wales, Sydney 2052, Australia; j.a.horner@unsw.edu.au
[2] Astronomy Group, Faculty of Social and Natural Sciences, Kinki University, Shinkamikosaka 228-3, Higashiosaka-shi, Osaka, 577-0813, Japan
[3] Research School of Astronomy and Astrophysics, and Planetary Science Institute, the Australian National University, ACT 0200, Australia

## ABSTRACT

The recent discovery of the first Neptune Trojan at the planet's trailing (L5) Lagrange point, 2008 LC18, offers an opportunity to confirm the formation mechanism of a member of this important tracer population for the Solar System's dynamical history. We tested the stability of 2008 LC18's orbit through a detailed dynamical study, using test particles spread across the ±3σ range of orbital uncertainties in *a, e, i* and *Ω*. This showed that the wide uncertainties of the published orbit span regions of both extreme dynamical instability, with lifetimes < 100 Myr, and with significant stability (> 1 Gyr lifetimes). The stability of 2008 LC18's clones is greatly dependent on their semi-major axis and only weakly correlated with their orbital eccentricity. Test particles on orbits with an initial semi-major axis of less than 29.91 AU have dynamical half-lives shorter than 100 Myr; in contrast, particles with an initial semi-major axis of greater than 29.91 AU exhibit such strong dynamical stability that almost all are retained over the 1 Gyr of our simulations. More observations of this object are necessary to improve the orbit. If 2008 LC18 is in the unstable region, then our simulations imply that it is either a temporary Trojan capture, or a representative of a slowly decaying Trojan population (like its sibling the L4 Neptunian Trojan 2001 QR322), and that it may not be primordial. Alternatively, if the orbit falls into the larger, stable region, then 2008 LC18 is a primordial member of the highly stable and highly inclined component of the Neptune Trojan population, joining 2005 TN53 and 2007 VL305. We attempted to recover 2008 LC18 using the 2.3 m telescope at Siding Spring Observatory to provide this astrometry, but were unsuccessful due to the high stellar density of its current sky location near the galactic centre. The recovery of this object will require a telescope in the 8m class.

# 1 INTRODUCTION

The Neptune Trojans, which orbit within the Neptunian 1:1 mean-motion resonance, are the most recently discovered small-body population of the Solar System. The existence of such objects was long postulated (Mikkola & Innanen, 1992), but due to their faintness, the first Neptune Trojan was not discovered until 2003 (Chiang et al., 2003). That object, 2001 QR322, had the near-circular, low eccentricity orbit expected of an object that formed in its current orbit from material in a dynamically cold protoplanetary disk (Marzari & Scholl, 1998; Chiang & Lithwick, 2005). In the years that followed, five more Neptune Trojans were discovered librating around Neptune's leading (L4) Lagrange point. However, only one of these, 2004 UP10, shared the low orbital inclination of 2001 QR322. Two had moderate inclinations (2005 TO74, $i = 5.244°$, 2006 RJ103, $i = 8.161°$).

Most startlingly, two Trojans had inclinations above 20°: 2005 TN53 ($i = 24.962°$) and 2007 VL305 ($i = 28.085$). Finding two highly inclined objects in such a small sample of the population implies that the highly-inclined Neptune Trojan population might significantly outnumber their low inclination counterparts (Sheppard & Trujillo, 2006). This unexpected part of the population is key to understanding the origin and evolution of the whole population.

The current paradigm for planetary formation requires significant migration of the outer planets before they reach their current locations (Goldreich & Tremaine 1980; Fernandez & Ip 1984, Hahn & Malhotra, 1999). This migration is invoked to explain the detailed features of the small-body populations, including, for example, the orbital eccentricity and inclination of Pluto (Malhotra, 1995) and the distribution of the Jovian Trojans (Morbidelli et al., 2005; Lykawka & Horner 2010). We have investigated the influence of Neptune's migration on the formation of the Neptune Trojans in great detail: the best origin for the Trojan population is that they were captured from non-Trojan orbits during Neptune's migration (Lykawka et al., 2009, 2010; Lykawka & Horner, 2010; Lykawka et al., 2011). Nesvorny & Vokrouhlicky (2009) independently obtained similar results using a computationally unrelated dynamical model.

However, even Neptune Trojans on orbits that imply stability may not have been acquired in this primordial capture. 2001 QR322 displays significant dynamical instability (Horner & Lykawka, 2010a). The original Neptune Trojan population probably contained many unstable objects; it continues to contribute them to the Centaur population as the Trojans gradually dynamically decay (Horner & Lykawka, 2010b, 2010c). 2001 QR322 may also be an interloper, captured after the migration of the planets and only temporarily visiting the Neptunian Trojan cloud (a process discussed in general terms by Horner & Evans, 2006). However, the long dynamical lifetime of 2001 QR322 suggests that it might well be a long-term resident that just happens to not be perfectly dynamically stable.

Dynamical simulations have shown that the Neptunian L5 point can host stable Trojans equally as well as the L4 (Holman & Wisdom 1993; Dvorak et al. 2007; Lykawka et al., 2009). However, the detection of Trojans librating around Neptune's trailing (L5) Lagrange point is difficult because that point currently lies near the galactic centre (Sheppard & Trujillo, 2010). As a result, it took the dedicated large-telescope image subtraction survey of Sheppard & Trujillo (2010) before the first L5 Trojan, 2008 LC18[1], was found. Their observations place it on a highly inclined (~27.5°) orbit that is slightly more eccentric than those of the other known Neptune Trojans ($e \sim 0.08$).

In their discovery paper, Sheppard & Trujillo (2010) stated that the seven Neptunian Trojans known at that time are dynamically stable on timescales comparable to the age of the Solar System. However, a suite of preliminary test simulations which we performed in late 2010 suggested that the orbit of 2008 LC18, like that of 2001 QR322, is unstable: approximately half of the orbital clones of 2008 LC18 did not survive as Trojans after 4 Gyr of orbital integration (Lykawka et al. 2011). Since 2008 LC18 was only recently discovered, the arc over which observations have been made is short, and so there remain significant errors in the estimated "best-fit" orbit. However, the precision with which its orbit is known is sufficiently high for us to attempt to draw conclusions on its stability and origin. For comparison, the first long-term dynamical investigations of 2001 QR322 were carried out once the uncertainties in its semi-major axis (*a*) and eccentricity (*e*) (expressed as the ratio of a given orbital element's nominal value to its 1σ uncertainty) were approximately 0.1% (Chiang et al. 2003; Brasser et al. 2004), and the current accepted orbit for 2008 LC18 has uncertainties in *a-e* that are of that order of magnitude. We therefore began a detailed dynamical study of the long-term dynamical behaviour of 2008 LC18, in order to find its place within our current understanding of the Neptunian Trojans.

---

[1] A second L5 Trojan was announced on 28 July 2011 in MPEC 2011-O47, 2004 KV18: it has a surprisingly high eccentricity $e = 0.184$, and a moderately inclined orbit with $i = 13.609°$.

## 2 OBSERVING 2008 LC18

To obtain new astrometry to reduce the uncertainty in its orbit, we observed 2008 LC18 on two nights in August 2011 with the Australian National University's 2.3m telescope at Siding Spring Observatory, NSW, Australia. Our one- to one-and-a-half hour observations with the imaging camera on 1 and 3 August were broken into multiple 5-minute exposures. The data were taken in 1.3" seeing, good conditions for the site. The Trojan's location in one of the densest regions of the Milky Way made it extraordinarily hard to pick out the R ~ 23.3 magnitude object. Each individual 5 min exposure reached a depth of R ~ 22.0, preventing the use of blinking to confirm the 2.4" movement of 2008 LC18 over the hour.

To attempt detection, we coadded each night's images, stacking along the predicted motion vector of 2008 LC18. This in principle would allow us to reach R ~ 23.3 to 5σ, assuming both accurate subtraction of the non-moving objects and accurate stacking on the known motion of the target. The target would be partly be subtracted during this process due to its small sky motion.

A deep image from the first night was used as the reference frame. The individual images were astrometrically aligned with *wcsremap*[2], then photometrically aligned by PSF-matching with *HOTPANTS*[3], and subtracted from the reference frame. The residual images were shifted to match the motion of the Trojan and coadded in *IRAF*[4]. No object was visible within the 1σ position uncertainty in the coadded frame. 2008 LC18 was also not visible when the coadded frames were blinked.

We conclude that deeper observations with a larger-aperture telescope in a location with better seeing are necessary to provide new astrometry.

## 3 THE SIMULATIONS

Following the method used for our in-depth study of 2001 QR322 (Horner & Lykawka, 2010a), we placed clones of 2008 LC18 around its nominal orbit in *a, e, i* and *Ω* parameter space, so that the clones spanned a region of ±3σ either side of the nominal value (Table 1) in each element.

|   | Value | 1σ variation | Units |
|---|---|---|---|
| a | 29.9369 | 0.02588 | AU |
| e | 0.083795 | 0.002654 |  |
| i | 27.569 | 0.003824 | deg |
| Ω | 88.521 | 0.0007854 | deg |
| ω | 5.135 | 10.85 | deg |
| M | 173.909 | 12.83 | deg |

**Table 1: The orbital elements of 2008 LC18 at epoch JD2455800.0, from the AstDys website, http://hamilton.dm.unipi.it/astdys/ on 23rd August 2011. Note the particularly large errors in the values of *ω* and *M* that hold the bulk of the uncertainty in the orbit of this object.**

---

[2] *wcsremap*: http://www.astro.washington.edu/users/becker/wcsremap.html
[3] *HOTPANTS*: HIGH ORDER TRANSFORM OF PSF AND TEMPLATE SUBTRACTION (Becker et al., 2004), http://www.astro.washington.edu/users/becker/hotpants.html
[4] *IRAF*: Image Reduction and Analysis Facility (Tody, 1993)

Given the great uncertainties in the values of $\omega$ and $M$, these orbital elements remained unchanged in all cases. In total, this gave us a test population of 61875 particles (25x15x15x11 in $a$-$e$-$i$-$\Omega$) which were then followed under the gravitational influence of Neptune, Uranus, Saturn and Jupiter for a period of 1 Gyr using the *N*-body dynamics package *MERCURY* (Chambers, 1999). Particles that collided with one of the massive objects were removed from the integration, and any object which reached a heliocentric distance of 1000 AU was taken as having been ejected from the Solar System, and was not followed further. Simulations of this duration are more than sufficient to allow us to obtain a good statistical view of the behaviour of the object, and are short enough to allow us to improve our statistics through the simulation of as many test particles as possible.

## 4 RESULTS AND DISCUSSION

Of the initial swarm of 61875 test particles, 18461 were either ejected from the Solar System, or collided with a giant planet, before the end of the 1 Gyr integration. 43414 of the test particles remained within the Solar System, a survival fraction of 70.2%. Following the procedure detailed in Horner & Lykawka (2010a), these results correspond to a dynamical half-life for the complete sample of approximately 1.96 Gyr (as compared to 593 Myr for 2001 QR322). The survival half-life is a proxy for the degree of stability of the objects as members of the Trojan cloud over long dynamical timescales. This suggests that 2008 LC18 is a dynamically unstable object, albeit on timescales a factor of ~3 longer than those of 2001 QR322[5].

However, given the size of the errors on the orbit for 2008 LC18, such instability might be the result of the wide spread of clones, rather than any inherent instability of the object itself. In addition, the decay curve for the population of 2008 LC18's clones is not well approximated by a simple exponential decay (though such a fit does give a reasonable first impression of the object's stability). Figure 1 shows the decay as a function of time, with the number of test particles remaining within the Solar System shown in black. The decay can be broken into two components: first, ~25% of the population decays rapidly, following an exponential decay with half-life of order <50-100 Myr. Once the bulk of this population is depleted, after ~200 Myr, the population decays much more slowly, as the remaining 75% of the population has a much longer dynamical lifetime.

---

[5] We note here that the dynamical lifetime of 2008 LC18 as a Solar system object will naturally be slightly longer than that of the object as a Neptunian Trojan – once clones escape from the Neptunian Trojan clouds, they become Centaurs (e.g. Horner & Lykawka, 2010b, c). The Centaurs are well known to be dynamically unstable on timescales of millions or tens of millions of years (e.g. Horner et al., 2004a, b), and so the great majority of escapees will survive as Solar system objects for several million years after their escape from the Trojan cloud. In Horner & Lykawka, 2010a, we calculated both the Trojan and Solar system lifetimes for 2001 QR322, and found them to differ by less than 10% (553 Myr vs. 593 Myr). For simplicity, in this work, we present solely the total lifetime of the object, with the implicit understanding that the lifetime *as a Trojan* would be slightly shorter. This is illustrated neatly by the fact that, of the 43414 test particles that remained within the Solar system at the end of our simulations, 42935 remained as Trojans, with just 479 having escaped that region but not yet been removed from the Solar system. This corresponds to a dynamical half-life *as a Trojan* for 2008 LC18, based on our results, of 1.90 Gyr, only slightly smaller than the 1.96 Gyr half-life as a Solar system object.

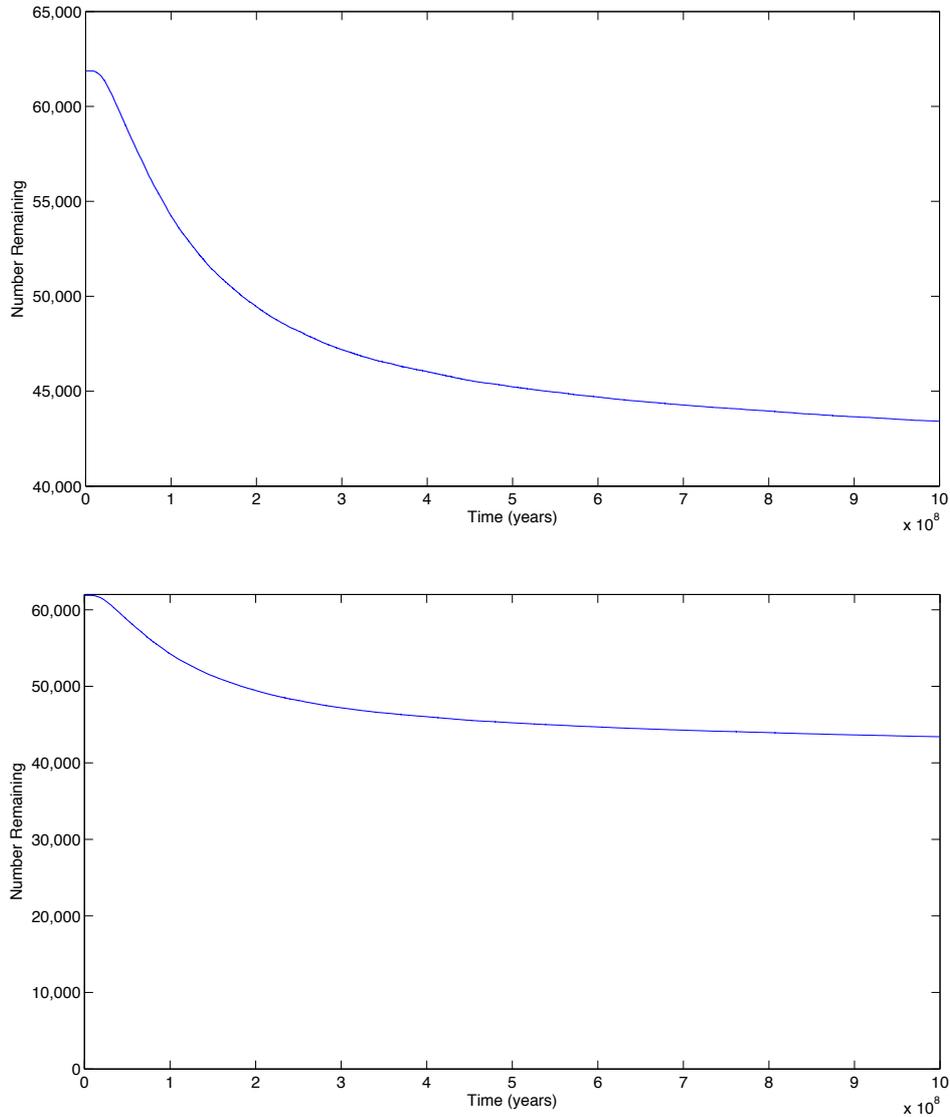

*Figure 1: The decay of the population of clones of 2008 LC18 as a function of time. Whilst both plots show the total number of test particles remaining, that in the upper panel is zoomed to show the decay in more detail. The decay of the population is best fit by the combination of two distinct subsets, each following their own distinct decay behaviour. Just over 25% of the test particles (the unstable component) decay on very short timescales, with an approximate decay half-life of order 50-100 Myr. Once those unstable objects have been depleted, the overall population decays far more slowly, indicative of the 75% of the test particles that move on orbits of significantly greater dynamical stability.*

As 2001 QR322 exhibited a strong dependence on its semi-major axis, with weak or no dependence on the other orbital elements (Horner & Lykawka, 2010a), we closely examined the variation in the stability of 2008 LC18 as a function of its orbital elements. For this we slightly oversampled the semi-major axis domain (25 clones in *a*, compared to 15 or 11 clones in the other elements).

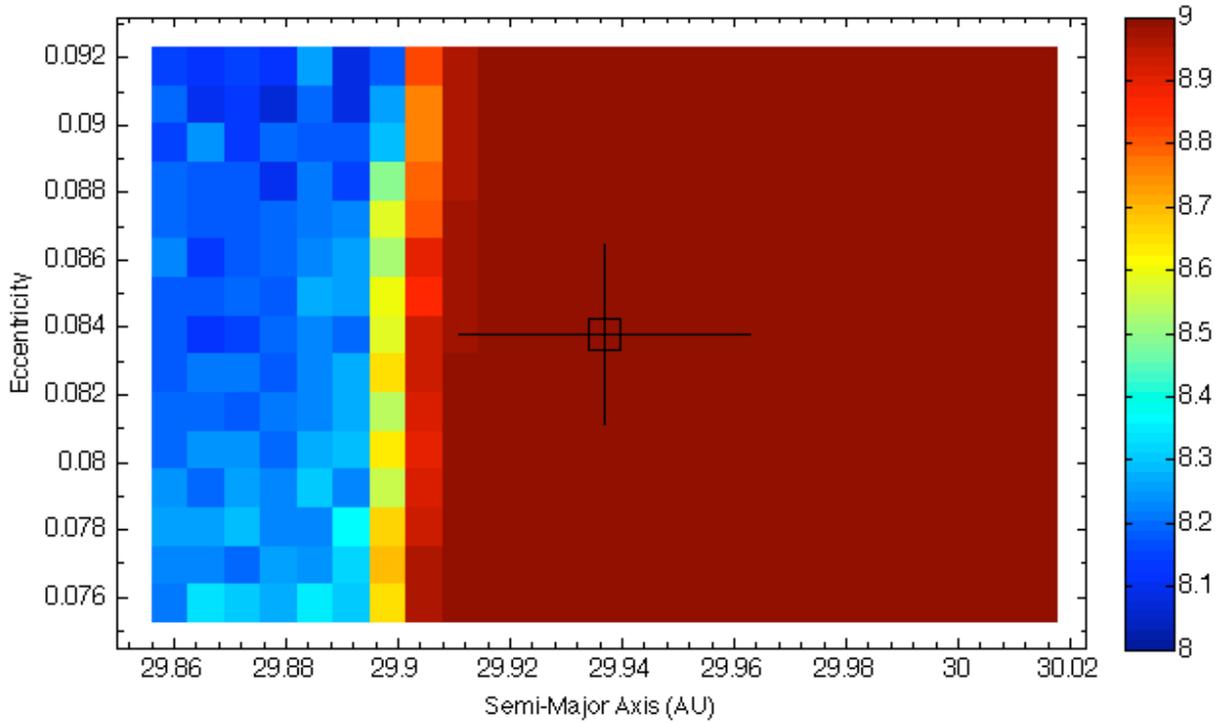

*Figure 2: The orbital stability of clones of 2008 LC18 as a function of their initial semi-major axis and eccentricity. The displayed orbital elements refer to a ±3σ distribution of particles, centred on the nominal orbit of the object at epoch JD2455800.0 (as detailed in Table 1), which is marked by the location of the box. The horizontal and vertical lines extending from the box show the ±1σ errors. Each box details the mean lifetime of a sample of 165 test particles which were followed for up to 1 Gyr. The lifetime of an individual test particle was deemed as ending at the instant it reached a distance of 1000 AU from the Sun or impacted upon one of the massive bodies in the simulated system (the Sun, Jupiter, Saturn, Uranus and Neptune). The test particles that survived to the end of the simulation were attributed a lifetime of 1 Gyr. Note the sharp division between very dynamically unstable orbits (blues, greens and yellows, left of ~29.91 AU) and strongly stable orbits (reds, to the right of that location)*

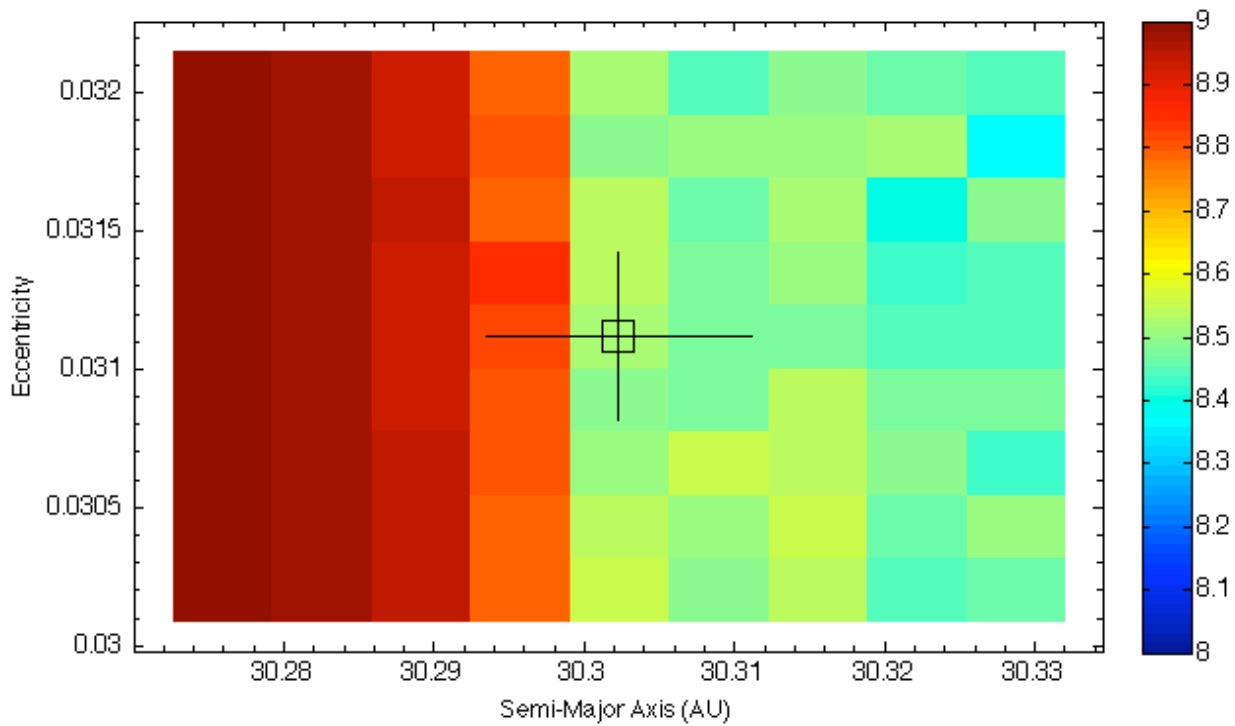

*Figure 3: The orbital stability of clones of 2001 QR322 as a function of their initial semi-major axis and eccentricity, for comparison to the behaviour presented in Figure 2 for 2008 LC18. The displayed orbital*

*elements refer to a ±3σ distribution of particles, centred on the nominal orbit of the object at epoch JD2454800.5 (as detailed in Horner & Lykawka, 2010a), which is marked by the location of the box. The horizontal and vertical lines extending from the box show the ±1σ errors. Each box details the mean lifetime of a sample of 243 test particles which were followed for up to 1 Gyr. The lifetime of an individual test particle was deemed as ending at the instant it reached a distance of 1000 AU from the Sun or impacted upon one of the massive bodies in the simulated system (namely the Sun, Jupiter, Saturn, Uranus and Neptune). The test particles that survived to the end of the simulation were attributed a lifetime of 1 Gyr. Note the sharp division between very dynamically unstable orbits (greens, left of ~30.33 AU) and strongly stable orbits (reds, to the left of that location).*

Figure 2 illustrates the relationship between the mean dynamical lifetime of the clones in initial semi-major axis vs. eccentricity element space. The mean lifetime of each region of the initial *a-e* space was calculated based on the 165 clones that began life in that region, giving us a total of 375 regions, each of which contained 165 clones (a total of 61875 clones).

The stability of the orbit of 2008 LC18 exhibits a remarkably strong dependence on the assumed initial semi-major axis. The role played by *e, i* and *Ω*, is negligible. This result is similar to 2001 QR322 (for comparison, in Figure 3), although in this case the distinction between the stable and unstable regions is far more pronounced. It is interesting to note that the area of instability of the orbit of 2008 LC18 is not simply restricted to the extreme regions of the ±3 sigma uncertainty range. Indeed, the unstable orbits begin at semi-major axes almost exactly one sigma from the nominal orbit. This, coupled with the structure of the unstable region (confined tightly in semi-major axis space) suggests that the potentially unstable behaviour we have uncovered might well be a fair representation of the orbit of 2008 LC18, rather than being a simple causal result of the large uncertainties in its best-fit orbit.

Moving beyond the simple relationship between the semi-major axis of the clones and their stability, it is also interesting to examine whether there is any relationship between their relative (in)stability and their libration amplitudes (the scale of the full angular motion of the object around the Lagrange point), *A*. When the libration amplitudes are determined for a sample of object in the dynamically stable region, we find that they move on orbits with typically $A < 50°$. By contrast, the objects that move in the unstable region, with typical lifetimes less than 300 Myr, display markedly larger initial libration amplitudes, typically in the range $A = 60$-$70°$. This is perhaps unsurprising, given that diffusion maps for orbits around the Neptunian Lagrange points (such as those produced by Nesvorny & Dones 2002; Marzari et al. 2003; Zhou et al. 2009) show that the region of stability for Trojans of inclinations similar to that of 2008 LC18 (i.e. $i \sim 27.5°$) seems to be constrained to values of $A < 60°$. In particular, it is interesting to note that the orbits of our unstable clones (moving with $i \sim 27.5°$ and $A = 60$-$70°$) seem to be located in the boundary between regular ("stable") and irregular ("unstable") orbits in the phase space detailed in the detailed diffusion maps of Zhou et al. (2009, 2010) (the regions shaded in purple in their fig. 11).

Given previous work on the (in)stability of objects in the Solar system's stable reservoirs, it seems reasonable to ask whether the region of instability observed is the result of the influence of any strong secular resonance between the Trojans and Neptune or another giant planet. Close inspection of the orbital and secular evolution of the bulk of the unstable clones revealed neither systematic large changes in the orbital eccentricity and inclination of the objects, nor any significant secular trends. In addition, we found no strong secular resonances involving the giant planets for these objects (such as first order apsidal or nodal secular resonances involving a planet, which would yield macroscopic features or trends in the behaviour of the orbits of these Trojans). However, it is worth noting that the orbits of the clones experiencing dynamical instability possess inclinations and libration amplitudes that seem to place them in a region close to the location of a web of complex secondary resonances. Those resonances involve the near 2:1 mean-motion resonance between Uranus and Neptune, the librational motion of the Trojan, and the apsidal motions of both Saturn and that of the Trojan (see Zhou, Dvorak & Sun 2009, 2011 for more details). This suggests that a family of such resonances may be the origin of the instability. Unfortunately, as a result of the relatively large uncertainties in the orbit of 2008 LC18 (which act to significantly hinder this type of analysis), we were unable to precisely determine the influence of this family of resonances on the objects. It is interesting to note that the destabilising role of such secondary resonances has already been found to play a significant role in determining the dynamical (in)stability of 2001 QR322 (as described in Horner & Lykawka 2010a). In that work, the authors highlight the fact that a secondary resonance of that family is the principle source of the instability towards the high-$a$ region of that object's orbital error ellipse. Taken together, these results highlight the rich and complex dynamics of the Neptune Trojans, features of which are clearly seen in dynamical maps of the Trojan's orbital element phase space.

The true nature of the object, then, seems strongly dependent on the orbit it currently occupies. As future observations refine and constrain the orbit ever more tightly, it will become apparent in which region it lies. Should the orbit be found to lie in the stable region, then it is reasonable to assume that, despite its high orbital inclination, 2008 LC18 is a primordial Neptunian Trojan, and has been librating around that planet's L5 Lagrange point since its formation or capture during the youth of the Solar System, as modelled by e.g. Lykawka et al., (2009, 2011). On the other hand, if the orbit is found to lie within the unstable region, it would in many ways be significantly more interesting.

In such a scenario, two possibilities spring out. Firstly, it might be the case that 2008 LC18 is a primordial Neptune Trojan that has recently begun to evolve from the Neptunian Trojan cloud to a more unstable orbit, through a process of slow diffusion. Such diffusion can simply be the result of the natural "random walk" of the object's orbital elements over time, as it responds to the distant perturbations of the Solar system's massive objects and the chaotic web of various high order resonances generated by those massive objects. Alternatively, the diffusion could also in part potentially be driven by mutual interactions between the larger members of the Neptunian Trojan population. Whilst this latter possibility is intriguing, we note that studies of the effect of the massive members of the non-resonant trans-Neptunian population (which are an order of magnitude larger in radius, and therefore significantly more massive than the Neptunian Trojans) have shown that they play only a negligible role in scattering their brethren (e.g. Lykawka et al., 2012). On the other hand, it might well be the case that such interactions play a more significant role when it comes to the destabilisation of objects trapped in mean-motion resonances. Indeed, Nesvorný et al. (2000) showed that the gravitational influence of Pluto can play a significant role in causing the excitation and eventual ejection of its fellow Plutinos from their residence in the Neptunian 2:3 mean-motion resonance. However, it should be noted once again that the mass of Pluto is far greater than that of any of the known Neptunian Trojans, and so the strength of their mutual gravitational perturbations would be proportionally smaller.

In addition to the effect of mutual gravitational perturbations, it is also possible that mutual collisions between members of the Trojan population could gradually excite their libration sufficiently for them to migrate from a stable to an unstable regime – particularly given that such a shift might only require the libration amplitude of a given Trojan to increase by of order ten degrees on Gyr timescales. Whilst it would be fascinating to study such interactions in more detail, to see whether they could play a role in the gradual dispersal of the Neptunian Trojans, such a study is beyond the scope of this work.

Regardless of the cause, such a diffusional origin for 2008 LC18's potential instability would place 2008 LC18 in the same category as 2001 QR322, and would add further weight to the conclusions of Horner & Lykawka (2010b, c) that the Neptune Trojans could represent a significant source of material to the Centaur population[6]. Since the Centaurs are the proximate source of the short-period comet population (Horner et al., 2003, 2004a, 2004b, 2009; Tiscareno & Malhotra 2003; di Sisto & Brunini 2007) the Neptune Trojans could well be a significant source of material to Earth-crossing orbits.

Given the extreme instability of clones in that region, however, a second possible solution would be that the object is not actually a primordial member of the Neptunian Trojan population, but rather has been captured in the relatively recent past, and is merely experiencing a temporary phase of Trojan behaviour. Such temporary captures have been observed for the planet Jupiter, which has captured a number of comets as temporary satellites over the past century (Ohtsuka et al. 2008; Kary & Dones 1996; Tancredi et al. 1990; Rickman & Malmort 1981). Trojan captures in particular have been seen in dynamical simulations of the outer Solar System (e.g. Horner & Evans, 2006), and have also been proposed as potential capture mechanism for the irregular satellites of the giant planets during the early stages of Solar system evolution (e.g. Koch & Hansen, 2011). However, we note that the longevity of the unstable clones as Neptune Trojans (timescales of tens of millions of years) are far longer than those recorded for the temporary Trojan captures around the other giant planets detailed in Horner & Evans (2006). It is therefore somewhat difficult to envision a scenario that would allow for such a long-term capture, without some non-gravitational damping mechanism to help trap the object in Neptune's Trojan cloud. We also note that, in that work, the authors failed to find a single occurrence of a temporary Neptunian Trojan capture, although this is most likely the result of the short duration of their integrations (just 3 Myr) and small population of test particles (just over 22,000). On the other hand, if it is confirmed that 2008 LC18 is truly a recently captured object, it will represent the first observed temporary capture of an object by Neptune, and may represent a modern-day counterpart to the process by which the Neptunian Trojans were captured in the first place.

---

[6] A more detailed discussion on the importance and implications of lost Trojan populations can be found in Lykawka & Horner (2010) and Horner & Lykawka (2010a).

Given the potential instability of this object, and the wide variety of outcomes that its current orbital uncertainties allow, we look forward to further observations being made to improve the precision of its orbit. Indeed, our work highlights the importance of obtaining better constrained orbits, or undertaking detailed dynamical analysis, before newly discovered objects are definitively categorised as members of the Solar system's resonant populations. Despite our best efforts to remedy the significant orbital uncertainties for 2008 LC18, we were unable to recover the object using the Australian National University's 2.3m telescope at Siding Spring Observatory, NSW, Australia, due to a combination of the objects faintness and the great degree of crowding resulting from the fact that 2008 LC18 is currently moving through one of the densest fields of the Milky Way. The recovery of this object would certainly be feasible with larger aperture telescopes cited at a location with better seeing, and we eagerly await such results.

Beyond simply recovering 2008 LC18 observationally, it would be interesting to see whether the colour and thermal properties of the object (as can be determined using the *HERSCHEL* space telescope (Müller et al., 2009, 2010) are similar to those of the other Neptune Trojans – noticeably different properties would provide independent evidence that the object is just a temporary visitor to the Neptunian Trojan cloud, while results showing that 2008 LC18 is similar to the other Neptune Trojans would add weight to the idea that it is a primordial resident of that population. Such conclusions will no doubt be strengthened once the JWST comes online, allowing the measurement of the deuteration of the Solar System's icy bodies (Lunine et al., 2010). Such measurements will provide definitive observation tests of the formation regions of those bodies (as discussed by Horner et al. (2007); Horner et al. (2008)), hopefully providing a final answer to the true origin of the Neptunian Trojans.

**ACKNOWLEDGEMENTS**
JH acknowledges the financial support of the Australian Research Council, through the ARC discovery grant DP774000. MTB is supported by an Australian Postgraduate Award and the Joan Duffield Research Scholarship. The authors wish to thank Ramon Brasser and an anonymous referee, who made a number of helpful comments and suggestions on how the manuscript could be improved.